\documentclass[10pt]{article}

\usepackage{amssymb}
\usepackage{amsmath}
\usepackage{bm}
\usepackage{hyperref}
\numberwithin{equation}{section}
\usepackage{pigpen}
\usepackage{lifecon}
\usepackage{color}

\begin{document}

\title{Time and the Algebraic Theory of Moments.}
\author{B. J. Hiley\footnote{E-mail address b.hiley@bbk.ac.uk.}.}
\date{TPRU, Birkbeck, University of London, Malet Street, \\London WC1E 7HX. }
\maketitle

\begin{abstract}
We introduce the notion of an extended moment in time, the duron.  This is  a region of temporal ambiguity which arises naturally in the nature of process which we take to be basic.  We introduce an algebra of process and show how it is related to, but different from, the monoidal category introduced by Abramsky and Coecke.  By considering the limit as the duration of the moment approaches the infinitesimal, we obtain a pair of dynamical equations, one expressed in terms of a commutator and the other which is expressed in terms of an anti-commutator. These  two coupled real equations are equivalent to the Schr\"{o}dinger equation and its dual.

We then construct a bi-algebra, which allows us to make contact with the thermal quantum field theory introduced by Umezawa.  This allows us to link quantum mechanics with thermodynamics.  This approach leads to  two types of time, one is Schr\"{o}dinger time, the other is an irreversible time that can be associated with a movement between inequivalent vacuum states.  Finally we discuss the relation between our process algebra and the thermodynamic origin of time.

\end{abstract}

\section{Introduction.}

In this paper we address the question of time in quantum mechanics.  The first and more commonly chosen option is  to treat time as an external parameter as one does in the Schr\"{o}dinger and Heisenberg equations of motion.  In the relativistic domain time is treated as the fourth component of a four-vector.  In non-relativistic quantum mechanics, the three space components are regarded as operators, why keep time as a parameter?  Surely it should be treated as an operator. However the attempt to treat time as an operator is regarded as  a failure for the reasons discussed by Pauli \cite{wp58} in his seminal paper on this topic.  As a caveat, we should point out that  recently there have been two papers \cite{eg02a}, \cite{eg02b} that have challenged this conclusion. 

Rather than following this line of argument I want to make a radical departure and consider both space and time as arising from a deeper level in which process is taken as fundamental \cite{db65}.  In earlier work along these lines Bohm, Hiley and Stuart \cite{dbbh70}, and Hiley \cite{bjh91} proposed that this underlying process be referred to as the {\em holomovement}.  In contrast to the present world view which has it roots in Democritus with its `atoms' having a set of preassigned properties, we want to explore a world envisaged by Heraclitus, where all is change, all is flux.  This will lead us, in the first approximation, to introduce two types of time, an unfolding (Schr\"{o}dinger) time, together with a moment or duron, a `now', which allows us to consider the precise time to be ambiguous within some interval $\Delta T =t_2-t_1$.  We will denote this ambiguous moment by $[T_1, T_2]$ where $T_1 (T_2) $ are some suitable elements of an algebra that are functions of $t_1 (t_2)$ respectively  \cite{bhmf97}.  

Using these ideas we will construct an algebra of moments, which we detail in section \ref{sec:AP}.  In such a structure we cannot attach a meaning to an instant or a sharp `point' in time except through some limiting procedure.  We will show that in this limit, we can recapture the quantum formalism in algebraic form.  By this we mean we recapture the quantum formalism in what is generally known as Heisenberg's matrix mechanics.  But in order to follow such a line of reasoning we must first address some very basic questions.

The first of these questions is to ask, ``What is a quantum object?"  The answer is surely obvious? An electron is a point-like quantum object!  Those simple words hide a perplexing riddle that takes us far from the comfort of our classical world.  Let us venture into this {\em quantum world} and illustrate the problem with a simplistic example originally proposed by Weyl. In this `toy' world, let us represent `shape' and `colour' as quantum operators that do not commute.  To make this world even simpler suppose there are only two shapes, sphere and cube, that are the `eigenvalues' of the `shape' operator and only two colours that are `eigenvalues' of the `colour' operatorÑ red, and blue.

We require to collect together an ensemble of red spheres\footnote{This example will be appreciated by cricketers everywhere.}.  In this world we must use one instrument to measure colour (e.g. a pair of spectacles that enables us to distinguish colours) and another incompatible instrument to measure shape. I decide first to collect together spheres and discard all the cubes.  I then decide to collect together those spheres that the colour-measuring device classifies as red.  I am done.  I have an ensemble of red spheres.  So what is the problem?  Just recheck that the objects in the ensemble are all in fact spheres.  We check by using the first pair of glasses again and find that half are now cubes! No permanent {\em either/or} in this world.  No permanent {\em and/and} either!  

Let us look closer and follow Eddington's \cite{ae58} suggestion that the elements of existence in the process world can be described by idempotents, $E^2=E$.   The eigenvalues, $\lambda_{e}$, of an idempotent is 1 or 0, existence or non-existence.  In symbols
\begin{eqnarray*}
E^{2} = E, \hspace{1cm} 	\mbox{with}  \hspace{1cm} 	\lambda_{e} = 1 \; \mbox{or}\; 0.
\end{eqnarray*}
If all idempotents commute,  existence is well defined.  However in quantum theory, idempotents do not always commute.  
\begin{eqnarray*}
[E_{a}, E_{b}] \neq 0	
\end{eqnarray*}
What then of existence?
\begin{eqnarray*}
\mbox{Either}\; E_{a} \;\mbox{or}\; E_{b},\hspace{0.5cm}\mbox{never}\hspace{0.5cm}		E_{a} \;\mbox{and}\; E_{b}	
\end{eqnarray*}
Existence, non-existence and in between?   Clearly no world of classical objects!  

What now is the position of reductionism?  It won't work because we cannot start with some set of basic building blocks with well defined properties.  We cannot separate objects into ensembles with well-defined properties.  How can we build stable structures if we cannot do that? And when the cube is blue, can we rely on it still being a cube as we try to build a structure of blue cubes?  

No structures at all? How can this be?  Quantum mechanics was introduced to explain stable structures.  Without quantum mechanics there is no stability of matter!  Without quantum mechanics there would be no atom as we know it;  no crystalline structures, no DNA, and no classical world.  But we see a 
the classical world.  We are the DNA unfolded in this world!  We probe quantum phenomena from our classical world, so naturally we insist on reductionism.  We strive to find the elementary objects, the atoms, the leptons, the baryons,  the quarks, the strings, the loops and the M-branes from which we try to reconstruct the world. 

Surely we are starting in the wrong place. Spencer-Brown \cite{gsb69} and Parker-Rhodes \cite{pr81} certainly thought so, so too did Lou Kauffman \cite{lk82}!  We should start with the {\em whole} process and then to make a description we must start from `distinction' or `difference'.  We start with a broad brush with which to make the initial differences.  We then find relations between these differences.  Within these preliminary differences we make finer distinctions and establish more relationships between these new differences. We then make yet more finer distinctions, establishing further relationships  and so on.  In this way we build a hierarchy of orders, to describe a structure process.

  Kaufman \cite{lk82} following Spencer-Brown \cite{gsb69} introduces the notion of `crossing' the boundary of a distinction, symbolised by $\lcroof{\textcolor{white}{G}}$ with  $\lcroof{\textcolor{white}{G}}\;\lcroof{\textcolor{white}{G}}=\lcroof{\textcolor{white}{G}}$ , an idempotent.   We will see in section \ref{sec:AA} that the distinction cross,  $\lcroof{\textcolor{white}{G}}$ , will be replaced by an idempotent in the algebra with which we will describe our structure-process.  Thus in symbolic form 
\begin{eqnarray*}
[T_1, T_2]\rightarrow \lcroof{\mbox {\large {\em T}}_1}\; {\mbox {\large {\em T}}}_2 \rightarrow \psi_L(t_1)E_a\psi_R(t_2)
\end{eqnarray*}
Here $E_a^2=E_a$, is some suitably chosen idempotent and $\psi_LE_a (E_a\psi_R)$ is an element of the left (right) ideal constructed using some suitably chosen idempotent, $E_a$.

  Notice we are not God-like looking {\em in}, but inside looking {\em out}.   Should we think of these distinctions as passive marks or are we going to allow for the fact that we are part of the process of making these distinctions? Are we participators?  Wholeness implies that we and our instruments are inside the whole process, yet our current theories start with the assumption that we and our instruments are outside our cosmos and so we  struggle to get back in!

At this stage we must pause.  The mere thought of ``putting ourselves back into it" traps us into thinking that there is something independent and separate to be put back in.  We should never have been {\em out of it} in the first place!  Now I hear alarm bells ringing.  ``He is going to suggest that we must put subjectivity back into our science whereas we know that the whole success of science has been to keep the subject out!"  That is true of classical physics, but quantum physics says we must at least put our measuring instruments back into the system.  Go further and ask ``Is nature basically subjective?"

As Bohr \cite{nb61} constantly reminds us, there is no separation between the system and its means of observation. He emphasises that this fundamental inseparability arises as a direct consequence of what he called ``the indivisibility of the quantum of action".  After warning us of the dangers of using phrases like `disturbing the phenomena by observation' and `creating physical attributes to atomic objects by measurements' he gives an even clearer statement of his position. He writes, ``I advocate the application of the word phenomenon exclusively to refer to the observations obtained under specified circumstances, {\em including an account of the whole experimental arrangement}" (my italics) \cite{nb61}. Because of the meaning Bohr attaches to the word `phenomenon', he insists that analysis into parts is {\em in principle} excluded. 

However Bohr himself as the observer, is still outside.  He claims to be a {\em detached} observer.  No pandering to subjectivity here.  But the question that fascinates me is ``How do we become detached?"  Let me spell out the problem.  I am assuming that the universe did come into being from some form of quantum fluctuation along the lines that is currently assumed.  The exact details as to whether this takes the form of a unique occurrence or in the form of a multiverse, or yet something else is of no significance for my argument here.  Any quantum birth must have evolved into our classical world and the question is what are the essential properties of this evolution for the emergence of a classical world to take place.

Bohm and I have already given a description of how this could happen in the context of the Bohm approach (Bohm and Hiley \cite{dbbh93}), but there we already start half-way along the road when we single out the particle and give it a `rock-like' status. However as we have argued earlier, the quantum particle is not `rock-like'.  It properties are not behaving as we would expect.  Instead we have a quasi-stable process many of whose properties are constantly transforming.  All that ultimately remains are the quasi-invariant processes,  the distinctions, the idempotents.   

Note the word `quasi-invariant' can be worrying.  Fine for the so called `elementary' particles like the muon, the $\Lambda$s, the $\Omega^-$ and so on, but surely some properties are immutable such as charge, baryon number, lepton number etc.  But even here when electron meets positron charge and lepton vanish, being left with a pair of photons.  Thus we are forced to look at nature from a very different perspective.  This perspective does not allow us to start with particles or even fields.  I do not need the popular story of decoherence to reach the classical world.  That is fine if a classical world already exists.  Then decoherence plays a vital role.  But we are using  another approach in which classical ideas are abstracted from the notion of an indivisible unity that was the baby universe.

\section{Activity and Process.	\label{sec:AP}}

I want to start from the flow of experiences we encounter from the time we leave our collective intellectual womb.  As Kauffman \cite{lk82} stresses, the primitive perception is {\em distinction}. We perceive differences, make distinctions and build an order. We do this through relationships.  We relate different differences. We perceive similarities in these differences and then look for the differences in these similarities and so on.  In this way we construct a hierarchy of order and structure in the manner detailed by Bohm \cite{db65} in his long forgotten paper {\em Space, Time and the Quantum Theory understood in Terms of Discrete structure Process}.

But the differences of what?  Just difference!  We experience a flux of sensations, which we must order if we are to make sense of our world.  We focus on the invariant features in that flux.  What is inside? What is outside? What is left? What is right? And so on.  More generally what is $A$, what is not-$A$.  But the distinction $A$/not-$A$ is not absolute in a world of process.  In a different flux of perceptions, $B$ and not-$B$ may become a distinction.  In this context it may not be possible to make the distinction between $A$ and not-$A$.  The processes are ontologically and epistemologically incompatible so that even distinction becomes a relative concept. Ultimately we could reach some domain when the distinction becomes absolute in that domain. Thus emerges the classical world with its absolute and stable distinctions.  But note that this ordering does not only apply to the material world.  It also applies to the world of thought.  Here it is quite clear that the observer, the $I$ of my mind, is part and parcel of the overall structure of the same mind.  It is here that we have direct experience with the notion of wholeness.  It is also here that we have direct experience of flux, activity and process philosophically highlighted by Schelling \cite{fs10} and Fichte \cite{jf94}.

But even here it is easy to slip back into the categories of objects being the primary, forgetting that these objects take their form from the very activity that is thinking.  I cannot capture this point better than Eddington \cite{ae58} when he wrote,
\begin{quote}
Causation bridges the gap in space and time, but the physical event at the seat of sensation (provisionally identified with an electrical disturbance of a neural terminal) is not the {\em cause} of the sensation; it {\em is} the sensation.  More precisely, the physical event is the structural concept of that which the sensation is the general concept.
\end{quote}
Or perhaps we should use the school of continental philosophers like Fichte \cite{jf94} who wrote, 

\begin{quote}

For the same reason, no real being, {\em no subsistence or continuing existence}, pertains to the intellect; for such being is the result of a process of interaction, and nothing yet exists or is assumed to be present with which the intellect could be posited to interact.  Idealism considers the intellect to be a kind of doing and absolutely nothing more.  One should not even call it an {\em active subject}, for such an appellation suggests the presence of something that continues to exist and in which an activity inheres.
\end{quote}

Idealism? Probably much too far for physicists, but the emphasis on activity {\em per se} and {\em not the activity of a thing} is the message to take.  Neither idealism nor scientific materialism, but something different.

How can we hope to begin a description of such a general scheme?   Start with Grassmannn \cite{hg95}.  In the process of thought we can ask the question ``Is the new thought distinct from the old thought, or is it one continuous and developing activity?  We find it easier to `hold' onto our description in terms of  the `old', $T_{1}$, and the `new', $T_{2}$.  But are they separate?  Clearly not!  The old thought has the potentiality of the new thought, while the new thought has the trace of the old thought.  They are aspects of one continuing process.  They take their form from the underlying process that is thought.  Each has a complex structure of yet more distinctions, so that each $T$ can be thought of as the tip of an `iceberg' of activity.

In order to symbolise this basic indivisibility, we follow Grassmann \cite{hg95} and Kauffman \cite{lk80}, \cite{lk87}  and enclose the relationship in a square bracket,  
$[T_{1}, T_{2}]$. Some properties of this bracket have already been discussed above.  Relationship is a start but not enough in itself. Our task then is to order these relationships into a multiplex of structure. To do that we need some rules on how to put these relationships together.

In my paper on {\em The Algebra of Process} \cite{bjh95} I tentatively suggested two rules of combination.  Firstly a multiplication rule, (3), that  defines a Brandt groupoid.  Secondly I introduce a rule for addition, (5).  These two binary relations taken together, of course, define an algebra.  Our defining relations are
\begin{eqnarray*}
(1)&	 [kA,kB] = k[A,B]	\hspace{0.4cm}			&\mbox{Strength of process}.\\
(2)&	[A,B]^* = - [B,A]				&\mbox{Process directed}.\\
(3)&	[A,B][B,C] = [A,C]	\hspace{1.1cm}			&\mbox{Order of succession}.\\
(4)&	[A,[B,C]] = [A,B,C] = [[A,B],C]		&\mbox{Associativity}.\\
(5)&	[A,B] + [C,D] = [A+C,B+D]			&\mbox{Order of coexistence}.
\end{eqnarray*}

Notice $[A,B][C,D]$ is NOT defined.

The importance of the groupoid in quantum theory has been pointed out by 
 Connes \cite{ac90}.  He recalled  Heisenberg's \cite{wh25}  original suggestion in which Weyl, also draws to our attention, namely,  that $x(t)$, the position of the electron in the atom, must be replaced by
\begin{eqnarray*}
X_{m n}(t)=\sum_{a}R_{mn}\exp[i(\nu_m-\nu_n)t].
\end{eqnarray*}
Notice how the `position, $X_{mn}(t)$' becomes a set of two-point objects, a set of transitions between energy eigenstates labelled by $m$ and $n$.  Thus, once again,  we are talking about transitions between one state and another, that is between structures defining {\em what has been to what will be}. 

When written in this form
the exponent ensures that the Ritz combination rule of atomic spectra can be satisfied, namely
\begin{eqnarray*}
\nu_{mj}+\nu_{jn}=\nu_{mn}.
\end{eqnarray*}
This result is needed when we form variables like $X_{nm}(t)^{2}$ which appear in the discussion of a quantum oscillator.  Heisenberg then proposed that the amplitudes combine as
\begin{eqnarray*}
R_{mn}=\sum_{j}R_{mj}R_{jn}.
\end{eqnarray*}
This was originally recognised as the rule for matrix multiplication. But, as Connes has pointed out, it is based on a more primitive structure, namely the groupoid.   Indeed it was a study of Heisenberg's original paper \cite{wh25} that led me originally to propose the relations (1) to (4) above, although at that time I was unaware that I was dealing with a recognised mathematical structure, a groupoid.

 I have shown elsewhere \cite{bjh95} \cite{bjh02a}  how the quaternions and indeed how a general orthogonal Clifford algebra emerges from the groupoid defined by the relations (1) to (4).  I don't want to present these ideas here again \cite{bjh11}.  Also a later summary of the main results of the emergence of orthogonal Clifford  algebras can be found in Hiley and Callaghan \cite{bhbc12}.   Rather
  I want to relate  the defining relations (1) to (5) above to a structure introduced by Kauffman \cite{lk80}, which he called the iterant algebra. 
   
To explain the ideas lying behind the iterant algebra, let us start with the plane and divide it into two, an `inside' and an `outside'.  Now introduce the activity of `crossing' the boundary \cite{gsb69}, $\lcroof{\textcolor{white}{I}}$ , and denote the activity of crossing from inside to outside by $[I, O]$, while the crossing from outside to inside is denoted by $[O, I]$. Here $I$ and $O$ are simply symbols denoting `inside' and `outside'.  This is the primary distinction.

Kauffman then generalises the notation and introduces a product defined by
\begin{eqnarray}
[A, B]\star[C, D]  =  [AC, BD]		\label{eq:KP}
\end{eqnarray}
and shows that one can also use this relationship to generate the quaternions.  Thus we have two structures with two different products producing the same algebra. But are they so different? When $B = C$ we have
\begin{eqnarray}
[A, B]\star[B, D] = [AB, BD] = B[A, D]	
\end{eqnarray}
In this way the products have been brought closer together.  In fact product (\ref{eq:KP}) above is simply an equivalence class of the Kauffman product. But notice product (\ref{eq:KP}) is undefined in our structure when  $B \neq C$ thus, in one sense, giving a more general structure.

 We have already suggested that we may write $[A , B]\rightarrow \lcroof{\mbox {\large {\em A}}}\; {\mbox {\large {\em B}}} \rightarrow A_LE_aB_R$.  An even more suggestive form is to complete the sequence   
 \begin{eqnarray}
  A_LE_aB_R\rightarrow A\rangle\langle B \rightarrow |a\rangle\langle b|.	\label{eq:KB}
 \end{eqnarray}
  Here $A$ and $B$ are elements of the algebra, while $a$ and $b$ are the eigenvalues of the  elements $A$ and $B$ regarded as operators in a Hilbert space.  In fact the symbol $\rangle$ was introduced by Dirac \cite{pd39} \cite{pd47} who called it the {\em standard} ket\footnote{The label $a$ was suppressed by Dirac leaving it understood provided no ambiguity arose.}.  It was introduced to prevent multiplication from the right, thus forming a left ideal.  To prevent multiplication from the left, a dual symbol $\langle\; $, ({\em standard} bra), was also introduced, this time forming a right ideal.  It should be clear that the joint symbol $\;\rangle\langle\;$ is playing a role analogous  to $E_a$, our idempotent. 
 \footnote{Symbolically we write $\rangle\langle\;\times\;\rangle\langle\;=\langle\rangle\;\rangle\langle\;$ with $\langle\rangle\;=1$.}
   Although Dirac called $\rangle$ a vector in Hilbert space, it has a more natural meaning in terms of an algebra as we we see in section \ref{sec:AA}.

 In order to stay on familiar territory we will use the last term in equation (\ref{eq:KB}).  With this identification we can relate our work to that of Zapatrin \cite{rz01} and of Raptis and Zapatrin \cite{irrz01}  who developed an approach through the incident algebra. In this structure the product rule is written in the form
\begin{eqnarray}
|A\rangle\langle B|\cdot|C\rangle\langle D| = |A\rangle\langle B|C\rangle\langle D| = \delta_{BC}|A\rangle\langle C|.		\label{eq:RZ}
\end{eqnarray}
Again this multiplication rule is essentially rule (3), the order of succession above.  But there is a major difference. When $B \neq C$ the product in equation (\ref{eq:RZ}) is zero, whereas we leave it undefined at this level.    

Finally we also want to draw attention to the work of Abramsky and Coecke \cite{sabc04} who argue that a process approach to quantum phenomena can best be described in terms of a symmetric moniodal category.  Our product (3) above is identical to the product used in this category.  There is a very close relationship between the algebraic structure we adopt in section \ref{sec:AA} and the diagrams used by Coecke \cite{bc05} which form part of a much more general planer algebra \cite{vj99}. 

However I will not discuss these relationships further here as I want to return to the basic ideas that are open to us when we look at process in terms of an algebra.

\section{The Intersection of the Past with the Future.}

We are focusing on process or flux,  via a notion of {\em becoming} which we symbolise by $[T_{1}, T_{2}]$. There will be many such relationships forming an ordered structure defining what we have called `pre-space' elsewhere.  (See Bohm \cite{db86}  and Hiley \cite{bjh91}.)   In other words these relationships are not to be thought of as occurring {\em in} space-time, but rather space-time is to be {\em abstracted from} this pre-space.  This is a radical suggestion so let me try to develop my thinking more slowly. 

Conventional physics is always assumed to unfold in space-time, the evolution being from point to point.  In other words physics always tries to talk about time development {\em at an instant}.  Any change always involves the limiting process \[\lim_{\Delta t\rightarrow 0}\frac{\Delta x}{\Delta t}.\]  But before taking the limit, it looks as if we were taking a point in the past $(x_{1}, t_{1})$ and relating it to a point in the future $(x_{2}, t_{2})$, i.e. relating what {\em was} to what {\em will be}.  But we try to hide the significance of this step by going to the limit $(t_{2} - t_{1}) \rightarrow 0$.  Then we interpret the change to take place at an instant, $t$.  Yet curiously the instant $t$ is a set of measure zero sandwiched between the infinity of that which has passed and the infinity of that which is not yet.  This is fine for evolution of point-like entities but is questionable when the evolution of extended structures is involved.

When we come to quantum mechanics, it is not positions that develop in time but wave functions, which like the Pauli spinor, can be treated as a special element of the algebra, namely, a minimal ideal in the algebra (see Hiley \cite{bjh02a}). Ideals are determined by idempotents and, as we have seen above, idempotents can be used as `separators'. But they are more than separators, they are the essence of the individual aspects of the process.

To clarify these notions let us recall Feynman's classic paper \cite{rf48} where he sets out his thinking that led to his `sum over paths' approach. There he starts by dividing space-time into two regions $R'$ and $R^{\prime\prime}$.  $R'$ consists of a region of space occupied by the wave function before time $t'$, while $R^{\prime\prime}$ is the region occupied by the wave function after time $t^{\prime\prime}$, with $t' < t^{\prime\prime}$.  Then he suggested that we should regard the wave function in region $R'$ as contain information coming from the `past', while the conjugate wave function in the region $R^{\prime\prime}$ representing information coming from the `future'\footnote{This is essentially the same idea that led to the notion of the anti-particle `going backwards in time', but here we are not considering `exotic' anti-matter.}.  The {\em possible present} is then the intersection between the two, which is simply represented by the transition probability amplitude $\langle\psi(R^{\prime\prime})|\psi(R')\rangle$.  From this Feynman derives the Schr\"{o}dinger equation.  But what I want to discuss here is $|\psi(R')\rangle\langle\psi(R^{\prime\prime})|$.  This is where all the action is!

Before taking up this point further, I would like to call attention to a similar notion introduced by Stuart Kauffman \cite{sk96} in his discussion of biological evolution.  Here it is clear that we are talking about an evolution of {\em structure}.  Kauffman discusses the evolution of biological structures from their present form into the {\em adjacent possible}. The adjacent possible contains only those forms that can develop from the immediate previous form.  Radical re-structuring is limited, small deformations are more likely.
This means that only certain sub-class of forms can develop out of the past.  Thus not only does the future form contain a trace of the past, but it is also constrained by what is `immediately' possible.  So any development is governed by the {\em tension between the persistence of the past, and an anticipation of the future} \cite{anw}.  

What I would now like to do is to build this notion into a dynamics. Somehow we have to relate the past to the future, not in a completely deterministic way, but in a way that constrains the possible future development.  The basic notion we need is thus structures which when represented in space-time cannot be localised. 
  Central to our structure is the `moment',  $[T_1,T_2]$.  In algebraic terms it is {\em a-local} but when represented in space-time is non-local; not only non-local in space, but also `non-local' in {\em time}.  It is a kind of `extension in time', a `duron';  a region of ambiguity where re-structuring is possible.  This ambiguity fits comfortably with the energy-time uncertainty principle.  Thus a process that involves energy changes cannot be described as unfolding at an instant except in some approximation.

\subsection{Bi-local Dynamics.}

 How then are we to discuss the dynamics of process, a dynamics which depends on this notion of a moment?  Let us  start in the simplest possible way by proposing that the basic dynamical function will involve two external times, giving rise to a {\em bi-local} model.  Thus we will discuss the time development of two-point functions of the form $[A(t_{1}), B(t_{2})]$.  We will show that we are led to a pair of equations (\ref{eq:HJMinus}) and (\ref{eq:HJPlus}) which depend on a mean time and a time difference.  We will then show that we capture the usual equations of motion in the limit $t_{1}\rightarrow t_{2}$.  We will then go on to exploit the bi-local structure.

Fortunately we do not have to start with quantum physics as we can motivate the idea entirely within classical physics.  Such functions are implicit in all variational principles that lie at the heart of modern physics. For example, in his classic work on optics, Hamilton \cite{rh67}, recognising the importance of Fermat's least-time principle, basically a principle involving two times.  He even suggested that both optics and classical mechanics could be united into a common formalism by introducing a two-point characteristic function, $\Omega(x_1, x_2)$.  Following on from Hamilton's work, Synge \cite{js60}, in his unique approach to general relativity, proposed that a two-point function, which he called the `world function', lies at the heart of general relativity\footnote{In modern parlance these functions are the generating functions of the symplectomorphisms in classical mechanics (see de Gosson \cite{mdg01}).}. Can we exploit these two-point functions to develop a new way of looking at dynamics?

Let us start by recalling that the use of the variational principle produces the classical Hamilton-Jacobi equation (see Goldstein \cite{g50}).  Specifically this emerges by considering a variation of the initial point $x_1$ of the trajectory.  Standard theory shows that by varying the initial point $x_1$, we can obtain the relations 
\begin{eqnarray}
\frac{\partial S}{\partial x_{1}}=p(x_{1})\hspace{1cm}
\frac{\partial S}{\partial t_{1}}+H_{1}=0.	\label{eq:HJ1}
\end{eqnarray}
where we have written $H_{1}=H\bigl[x_{1},\partial S(x_1,x_2)/\partial x_{1}\bigr]$  for convenience and we have replaced the world function $\Omega$ by the classical action function $S$.  

What is not so well known is that if we vary the final point $B$, we find another pair of equations
\begin{eqnarray}
\frac{\partial S}{\partial x_{2}}=-p(x_{2})\hspace{1cm}
\frac{\partial S}{\partial t_{2}}-H_{2}=0.	\label{eq:HJ2}
\end{eqnarray}
Here the second Hamilton-Jacobi equation formally becomes the same by writing $t_{2} = -t_{1}$.  

Similarly for the quantum propagator $K(x_2,x_1,t_2,t_1)$ which we write as $K(2,1)$ \cite{rfah65}, we find not only
\begin{eqnarray}
i\hbar \frac{\partial K(2,1)}{\partial t_1}+K(2,1)H_1=0,	\label{eq:prop1}
\end{eqnarray}
but also
\begin{eqnarray}
i\hbar \frac{\partial K(2,1)}{\partial t_2}-H_2K(2,1)=0.	\label{eq:prop2}
\end{eqnarray}
The similarity in form between equations (\ref{eq:HJ1}) and (\ref{eq:HJ2}) and the equations (\ref{eq:prop1}) and (\ref{eq:prop2}) is not coincidental, but arises from the lifting properties from the classical symplectic group to its covering group, the metaplectic group (see de Gosson \cite{mdg01}.)
Could this similarity be taken to support the idea that we have a wave coming from the `past' and the `future', thus fitting into the general scheme I am developing here?

Leaving that speculation aside, let us see how we can formally exploit the two Hamilton-Jacobi equations (\ref{eq:HJ1}) and (\ref{eq:HJ2}). Consider a pair of points with co-ordinates $(x_{1}, t_{1})$ and $(x_{2}, t_{2})$ joined by a geodesic in configuration space. The world function (generalised action) for this pair can be written as $S(x_{1}, x_{2}, t_{1}, t_{2})$. (See de Gosson \cite{mdg06} for a formal treatment of the above structure.)
 
We will find it more convenient to use `sums' and `differences' rather than the co-ordinates themselves.  Thus we change to co-ordinates $(X, \Delta x, T, \Delta t)$ where
\begin{eqnarray*}
X=\frac{x_{1}+x_{2}}{2},\quad T=\frac{t_{1}+t_{2}}{2},\quad\Delta x=x_{2}-x_{1},\quad\Delta t=t_{2}-t_{1}.
\end{eqnarray*}
so that the generalised action becomes
\begin{eqnarray*}
S(x_{1}, x_{2}, t_{1}, t_{2})=S(X, \Delta x,T,\Delta t)
\end{eqnarray*}
Then equations (\ref{eq:HJ1}) and (\ref{eq:HJ2}) can be replaced by
\begin{eqnarray}
\frac{\partial S}{\partial X}=\Delta p,\quad \frac{\partial S}{\partial T}=\bigl[H_{2}-H_{1}\bigr]	\label{eq:HJMinus}\\		
\frac{\partial S}{\partial \Delta x}=P,\quad \frac{\partial S}{\partial \Delta t}=\tfrac{1}{2}\bigl[H_{2}+H_{1}\bigr]\label{eq:HJPlus}	
\end{eqnarray}
In order to see the meaning of the two equations let us make a Legendre transformation
\begin{eqnarray}
K(X,P,T,E)=P\Delta x+E\Delta t -S(X, \Delta x,T,\Delta t)
\end{eqnarray}
so that
\begin{eqnarray*}
\frac{\partial S}{\partial T}=-\frac{\partial K}{\partial T},\quad
\frac{\partial S}{\partial\Delta t}=E
\end{eqnarray*}
A general background discussion to these ideas can be found in Bohm and Hiley \cite{dbbh81}.

Equations (\ref{eq:HJMinus}) and (\ref{eq:HJPlus}) will form the basis of a bi-local classical theory.  Now we must show that if we go to the limit $\Delta t\rightarrow 0$ and $\Delta x\rightarrow 0$ we will reproduce the expected equations of motion.  Therefore let us  go to this limit.  We find
\begin{eqnarray}
\lim_{\Delta t\rightarrow0}\frac{\partial S}{\partial T}=-\bigl[H_{2}-H_{1}\bigr]\Rightarrow\frac{\partial S}{\partial T}+\frac{\partial H}{\partial P}\Delta p +\frac{\partial H}{\partial P}\Delta x\approx 0.	\label{eq:LimHJ1}
\end{eqnarray}
But
\begin{eqnarray*}
\Delta p=-\frac{\partial K}{\partial X}\quad \Delta x=\frac{\partial K}{\partial P},
\end{eqnarray*}
so that equation (\ref{eq:LimHJ1}) becomes 
\begin{eqnarray}
\frac{\partial K}{\partial T}+\{K,H\}=0	\label{eq:CL}
\end{eqnarray}								
where $\{,\}$ is the Poisson bracket so that equation (\ref{eq:CL}) becomes the classical equation of motion for the dynamical variable $K$.  Indeed when $K$ is identified with the probability distribution, this is nothing more than the Liouville equation. 

The second equation in (\ref{eq:HJMinus}) becomes 
\begin{eqnarray*}
\frac{\partial S}{\partial \Delta t}=\tfrac{1}{2}\bigl[H_{2}+H_{1}\bigr]=E.	\end{eqnarray*}
Since $E$, the total energy, is a constant for a closed system, we have 
\begin{eqnarray}
\lim_{\Delta t \rightarrow 0}\frac{\partial S}{\partial \Delta t}   \label{eq:KE}=E.
\end{eqnarray}
Thus we see that in the limit $\Delta t\rightarrow 0$, the dynamics is defined by two equations, namely, equation (\ref{eq:CL}) and equation (\ref {eq:KE}).  They are both conservation equations, the first is the conservation of probability and the second is the conservation of energy.  We will now show that the analogue of these two conservation equations also emerge in the quantum case as we will now show.

\section{Quantum Pasts and Futures.}

\subsection{The Hilbert Space Approach.}

 Now let us examine the quantum domain and consider Feynman's suggestion mentioned earlier in more detail.  Introduce a world function defined by
 \begin{eqnarray}
 \hat\rho(t_{1},t_{2})=|\psi(t_{1})\rangle\langle\psi(t_{2})|.
 \end{eqnarray}								
 We use the symbol $\hat\rho$  because it will turn out that we are  essentially dealing with a generalised density operator.  Let us proceed formally by writing
 \begin{eqnarray}
 \frac{\partial }{\partial T}(|\psi(t_{1})\rangle\langle\psi(t_{2})|=
 \left( \frac{\partial }{\partial t_{1}}|\psi(t_{1})\rangle\right)\langle\psi(t_{2})|+|\psi(t_{1})\rangle\left( \frac{\partial }{\partial t_{2}}\langle\psi(t_{2})|\right)  \label{eq:QL1}
 \end{eqnarray}								
 We could use the two equations (\ref{eq:prop1}) and (\ref{eq:prop2}) in (\ref{eq:QL1}) to proceed, but since Feynman has already derived the Schr\"{o}dinger equation from these considerations, we prefer to substitute these two equations 
 \begin{eqnarray*}
 i\frac{\partial}{\partial t_{1}}|\psi(t_{1})\rangle=\hat H_{1}|\psi(t_{1})\rangle\quad\mbox{and}\quad -i\frac{\partial}{\partial t_{2}}\langle\psi(t_{2})|=\hat H_{2}\langle\psi(t_{2})|
 \end{eqnarray*}
 into equation (\ref{eq:QL1}) § to find   
 \begin{eqnarray}
 i\frac{\partial \hat\rho(t_{1},t_{2})}{\partial T}+\hat\rho(t_{1},t_{2})\hat H_{2}-\hat\rho(t_{1},t_{2})\hat H_{1}=0
 \end{eqnarray}								
 If we now take the limit as $\Delta t \rightarrow 0$ when $T \rightarrow t$, we find
 \begin{eqnarray}
 i\frac{\partial \hat\rho}{\partial t}+[\hat\rho,\hat H]_{_{-}}=0	\label{eq:QL}
 \end{eqnarray}								
 Here  $\hat\rho$ has become the usual density operator for the pure state $|\psi(t)\rangle$.  This equation is the quantum version of equation (\ref{eq:HJMinus}) and is, of course,  the quantum Liouville equation.

Now let us consider
\begin{eqnarray*}
2 \frac{\partial }{\partial \Delta t}\left(|\psi(t_{1})\rangle\langle\psi(t_{2})|\right)=|\psi(t_{1})\rangle\left( \frac{\partial }{\partial t_{2}}\langle\psi(t_{2})|\right)-
 \left( \frac{\partial }{\partial t_{1}}|\psi(t_{1})\rangle\right)\langle\psi(t_{2})|.
\end{eqnarray*}
So that by using the  two Schr\"{o}dinger equations again, we find this time
\begin{eqnarray}
2i\frac{\partial \hat\rho(t_{1},t_{2})}{\partial \Delta t}+\hat\rho(t_{1},t_{2})\hat H_{2}+\hat\rho(t_{1},t_{2})\hat H_{1}=0,	\label{eq:HJPlus1}
\end{eqnarray}								
which we recognise as the quantum version of equation (\ref{eq:HJPlus}).  The `derivative' $\partial/\partial\Delta t$ looks rather odd until one recalls field theory,
  \begin{eqnarray*}
\lim_{t_2\rightarrow t_1}\left[|\psi(t_{1})\rangle\left( \frac{\partial }{\partial t_{2}}\langle\psi(t_{2})|\right)-
 \left( \frac{\partial }{\partial t_{1}}|\psi(t_{1})\rangle\right)\langle\psi(t_{2})|\right]=|\psi(t_1)\rangle\overleftrightarrow{\partial_t}\langle\psi(t_2)|
 \end{eqnarray*}
With a little work, we can show
\begin{eqnarray}
|\psi(t_1)\rangle\overleftrightarrow{\partial_t}\langle\psi(t_2)|=T^{00}=E
\end{eqnarray}
 Thus we can finally write equation (\ref{eq:HJPlus1}) as
\begin{eqnarray}
2E=[\hat\rho,\hat H]_{_{+}}
\end{eqnarray}  \label{eq:QE}								
This turns out to be an expression of the conservation of energy equation.  
Collecting together the main results so far we find
\begin{eqnarray*}
 i\frac{\partial \hat\rho}{\partial t}+[\hat\rho,\hat H]_{_{-}}=0\quad\Leftrightarrow\quad  \frac{\partial K}{\partial T}+\{K,H\}=0
\end{eqnarray*}
and
\begin{eqnarray*}
2 E=[\hat\rho,\hat H]_{_{+}}\quad \Leftrightarrow \quad E=\lim_{\Delta t\rightarrow 0}\frac{\partial S}{\partial \Delta t}
\end{eqnarray*}
Again if $K$ is the classical analogue of the density operator then we would have a correspondence between the classical `Liouville' equation (\ref{eq:CL}) and the quantum Liouville equation (\ref{eq:QL}).  In turn the quantum energy equation (\ref{eq:QE}) then corresponds to the classical energy equation  (\ref{eq:KE}). Thus we have a clear correspondence between the classical and the quantum levels. 


\subsection{The Algebraic Approach.	\label{sec:AA}}

So far we have restricted our discussion to the more conventional mathematics, but I want to exploit a more general way of exploring these ideas using an algebraic approach that has already been discussed in  Brown and Hiley \cite{mbbh00} and further developed by Hiley \cite{bjh11} and Hiley and Callaghan \cite{bhbc12}.   

In the algebraic approach, a ket $|\psi(t_{1})\rangle$ is replaced by an element of a minimal left ideal, $\Psi_{L}(t_{1})$, while $\langle\psi(t_{2})|$ is replaced by an appropriate element of a right ideal, $\Psi_{R}(t_{2})$\footnote{If $R$ is a noncommutative ring, a left ideal is a subset $I_L$ such that if $a\in I_L$ then $ra\in I_L$ for all $r\in R$.}.

We then start by defining an algebraic density element 
\begin{eqnarray*}
\bar\rho(t_{1},t_{2})=\Psi_{L}(t_{1})\Psi_{R}(t_{2})
\end{eqnarray*}
and  write these algebraic elements, $\Psi$, in polar form
\begin{eqnarray*}
\Psi_{L}(t_{1})=R(t_{1})\exp[iS(t_{1})]\quad\mbox{and}\quad 
\Psi_{R}(t_{2})=\exp[iS(t_{2})]R(t_{2})
\end{eqnarray*}
Here we emphasise that $\Psi, R, S$ are {\em elements of the algebra} and not elements of a Hilbert space.  Then
\begin{eqnarray*}
2\frac{\partial \bar\rho(t_{1},t_{2})}{\partial \Delta t}=
\left[-\frac{\partial R(t_{1})}{\partial t_{1}}R(t_{2}) +R(t_{1})\frac{\partial R(t_{2})}{\partial t_{2}}
-i R(t_{1})R(t_{2})\left[\frac{\partial S(t_{1})}{\partial t_{1}}+\frac{\partial S(t_{2})}{\partial t_{2}}\right] \right]\nonumber\\
\times \exp \left(-i\left[S(t_{2})-S(t_{1})\right]\right)  \hspace{5cm}
\end{eqnarray*}									
where we have assumed that $R$  and $S$  commute.  Then when we go to the limit $\Delta t \rightarrow 0$ with $T \rightarrow t$, we find
\begin{eqnarray}
\lim_{\Delta t\rightarrow 0}2\frac{\partial\bar\rho}{\partial t}=-iR^{2}\frac{\partial S}{\partial t}
\end{eqnarray}								
Thus equation (\ref{eq:HJPlus1}) then become
\begin{eqnarray}
2R^{2}\frac{\partial S}{\partial t}+[\bar\rho,H]_{_{+}}=0	\label{eq:SE}
\end{eqnarray}								
This equation is identical to equation (11) derived in Brown and Hiley \cite{mbbh00}.  A yet different derivation of this equation will also be found in Hiley \cite{bjh02}.  The reason why I have re-derived this equation in different ways is because I have not seen this equation written down in this form in the literature. However it is implicit in Dahl \cite{jpd}

In Brown and Hiley \cite{mbbh00}  we showed that there were two important consequences following from this equation.  Firstly the Berry phase and the Aharonov-Bohm effect followed immediately from this equation in a very simple way.  Secondly we used this quantum equation to see where the quantum potential introduced by Bohm emerges from what is essentially the Heisenberg picture (see also Hiley \cite{bjh02}).  We found that this potential only appeared as a result of {\em projecting} the algebraic elements onto a representation space.  This led us to speculate that all the `action' of quantum phenomena takes place in a pre-space, the structure of which is described by the algebra.  All we see is its projection onto a space-time manifold.  Thus the space-time manifold is not to be taken as `basic'.  Rather it is something that is derived from the deeper and more basic structure-process.

It is well-known that we cannot display quantum processes in a {\em commutative} phase space because we are using a non-commutative structure. However this does not rule out the possibility of representing quantum phenomena in terms of a {\em non-commuting} phase space.  In fact this has already been achieved through the Moyal algebra \cite{jm49}, sometimes described as the deformed Poisson algebra. 

This structure contains a non-commutative $\star$-product which gives rise to a Moyal bracket, which can be used to produce an analogue of equation (\ref{eq:QL}).  There also exists a symmetric bracket, the Baker bracket  \cite{bjh04}, which can be used to produce an equation which is the analogue of  (\ref{eq:QE}).  Thus these equations seem basic to the type of non-commutative structures that we are using to describe quantum phenomena.  

What is even more interesting is that the Moyal algebra provides a natural way to approach the  classical limit.  The Moyal bracket equation reduces the classical Liouville equation which leads to a conservation of probability, while the  equation involving the Baker bracket reduces to the classical Hamilton-Jacobi equation.  We have shown the details elsewhere, \cite{mbbh00}, where we also show that when the quantum form of this equation is projected into a space representation, the quantum potential emerges through what we have called the quantum Hamilton-Jacobi equation.  We now see why this QHJ approaches the ordinary Hamilton-Jacobi equation in the classical limit.  The appearance of the quantum potential is clearly a consequence of the non-commutative structure required by quantum theory.

This leads to an interesting connection with the work of Gel'fand \cite{gl97} where it can be shown that for any {\em commutative} $C^*$-algebra, one can reconstruct the Hausdorff topological space $M$ underlying the commutative algebra. With a non-commutative algebra there is no unique underlying manifold.  One has to introduce a set of `shadow' manifolds, which are constructed by sets of projections from the algebra.  In each projection, we get a kind of distortion of the type found in maps when using  a Mercator's projection.  Therefore it is not surprising to find it necessary to introduce inertial forces, like the one derived from the quantum potential, to account for the predicted behaviour in the shadow manifold.  This is very similar to how the gravitation force is manifested in general relativity (For a more detailed discussion of these ideas, see Hiley \cite{bjh11}).

\section{Bi-Algebras and super-algebras.}

\subsection{Motivation.}

In this next section I want to extend the algebra and construct a bi-algebra.  This is motivated by some proposals made by Umezawa \cite{hu93}  in his discussions of thermal quantum field theory. His aim was to find a common formalism in which both quantum and thermal effects can be incorporated.  Unlike the work presented here, Umezawa uses Hilbert space and shows that if we `double' the Hilbert space, then the thermal state can also be represented by a single vector in this double space.  For example, in more familiar notation, the thermal wave function can be written in the form
\begin{eqnarray}
|\Omega(\beta)\rangle=Z^{-1/2}\sum\exp[-\beta E_{n}/2]|\psi_{n}\rangle\otimes|\psi_{n}\rangle.	\label{eq:U2H}
\end{eqnarray}								

Here $\beta =1/kT$   and $|\psi_{n}\rangle$ are the energy eigenkets.  $Z$ is the partition function.  The ensemble average of some quantum operator $A$ would then be given by
\begin{eqnarray*}
\langle\Omega(\beta)|A|\Omega(\beta)\rangle=Tr(\rho A).
\end{eqnarray*}
where $\rho$ is the thermal density operator, which in its  more usual form is written as
\begin{eqnarray*}
\rho=\exp[-H\beta]
\end{eqnarray*}

 Those familiar with algebraic quantum field theory will recognise that the doubling of Hilbert space is essentially the GNS construction (Emch \cite{ge72} and Hiley \cite{bjh02a}).  In terms of the algebra, this doubling of the number of field elements suggests that any algebraic theory would have double the algebra, but the bi-local theory I have introduced above is the first step to developing a bi-algebraic structure. 
 
In the last section, we have been discussing a two-time quantum theory where the time is being treated as a parameter and not as an element of a general non-commutative algebra.  Let us now see how we can generalise the structure to make time part of the larger algebra. 
 
 In order to anticipate the quantum approach, we return to classical physics and form a Poisson bi-algebra by introducing the  a generalised Poisson bracket defined by
\begin{eqnarray*}
\{\;\}=\frac{\partial }{\partial X}\frac{\partial }{\partial \Delta p}-\frac{\partial }{\partial \Delta p}\frac{\partial }{\partial X}+\frac{\partial }{\partial \Delta x}\frac{\partial }{\partial P}-\frac{\partial }{\partial P}\frac{\partial }{\partial \Delta x}.
\end{eqnarray*}
 so that we find the following relationships
 \begin{eqnarray}
 \{X,\Delta p\}=\{\Delta x,P\}=1\hspace{2cm}\nonumber\\
  \{X,P\}= \{\Delta x,\Delta p\}= \{X,\Delta x\}= \{P,\Delta p\}=0  \label{eq:DP}
 \end{eqnarray}								
This suggests we introduce another pair of brackets of the form
\begin{eqnarray}
\{T,(H(t_{2})-H(t_{1}))\}=\{\Delta t,(H(t_{2})+H(t_{1}))\}=1.	\label{eq:T,H}
\end{eqnarray}								
If we were to introduce the quantity $L(t_{1}, t_{2}) = H(t_{2}) - H(t_{1})$, we have the classical correspondence to the Liouville operator introduced by Prigogine \cite{ip80}.  This connection will be discussed further when these results are generalised to the quantum domain.  

\subsection{The Quantum Bi-Algebra.}

In moving to quantum theory, we need to regard the position and momentum as {\em algebraic elements} and base the theory on pairs of {\em algebraic} elements, $\{\bar x_{1},\bar x_{2}, \bar p_{1}$ and $\bar p_{2}\}$.  Again I have added the `bar' to emphasise that these are elements of the algebra.  In other words we are doubling  the algebra to form a bi-algebra.  Following the analogous procedure to the classical case, we introduce the notation
\begin{eqnarray}
2\bar X=\bar x_{1}\otimes 1+1\otimes \bar x_{2},\quad\bar\eta=\bar x_{1}\otimes 1-1\otimes \bar x_{2} \label{eq:BP1}\\						
2\bar P=\bar p_{1}\otimes 1+1\otimes \bar p_{2},\quad\bar\pi=\bar p_{1}\otimes 1-1\otimes \bar p_{2}	\label{eq:BP2}
\end{eqnarray}									
We then find that the following commutator relations hold
\begin{eqnarray*}
[\bar X, \bar\pi]=[\bar\eta,\bar P]=i
\end{eqnarray*}
and
\begin{eqnarray}
[\bar X, \bar P]=[\bar\eta,\bar\pi]=[\bar X,\bar\eta]=[\bar P, \bar\pi]=0 \label{eq:BC}
\end{eqnarray}									

These relations are the quantum analogues of the generalised Poisson brackets defined in equation (\ref{eq:DP}).  These results were already reported in Bohm and Hiley \cite{dbbh81}.  

The aim in this section of the paper is to find a time `operator' that may be connected with irreversibility.  Prigogine \cite{ip80} has already pointed out  that we need a theory in which irreversibility plays a fundamental role directly in the dynamics itself.  Let us see how we can make contact with his approach.  

First note that we can write the quantum Liouville equation (\ref{eq:QL}) in terms of the bi-algebra
\begin{eqnarray}
i\frac{\partial \bar\rho_{V}}{\partial t}+\bar L\bar\rho_{V}=0
\end{eqnarray}									
Here $\bar\rho_{V}$  is a vector equivalent of the density operator and $\bar L=\bar H\otimes 1- 1\otimes \bar H$.  The appearance of the `super-operator' $\bar L$ enables us to introduced a time `operator' $\bar T$ , defined through the relation 
\begin{eqnarray}
[\bar T,\bar L]=i
\end{eqnarray}									
This is the quantum version of the classical form presented by the first equation in (\ref{eq:T,H}). 

Prigogine \cite{ip80} argues that this time operator, $\bar T$  represents the `age' of the system. I don't want to discuss the reasons for this as I have already made some comments on it in Bohm and Hiley \cite{dbbh81}  and in Hiley and Fernandes, \cite{bhmf97}.  A more general discussion of Prigogine's point of view will be found in George and Prigogine \cite{gp79}, and in Prigogine \cite{ip80}, 

What I want to do now is to go on to the bi-algebraic generalisation of equation (\ref{eq:QE}).  This requires the introduction of the `super-operator' corresponding to the anti-commutator, which can be written in the form
\begin{eqnarray}
\bar E\bar\rho_V=(\bar H\otimes 1 +1 \otimes\bar H)\bar\rho_{V}=E_{+}\bar\rho_{V}.
\end{eqnarray}									
Such an operator was first introduced by George {\em et al} \cite{cgfh78} in their general discussion of dissipative processes.  They, like us, regard this as an expression of the total energy of the system. I have only found one other discussion relating  the anti-commutator, $[\bar\rho,H]_+$,  to the energy of the system.  This is the work of Dahl \cite{jpd} who was concerned with energy storage and transfer in chemical systems.

For completeness  I should point out that Fairlie and Manogue \cite{df91}  have discussed an analogous equation based on the cosine Moyal bracket introduced by Baker \cite{gb58}.  However they explore a very different structure.

As well as introducing  the `age operator', $\bar T$ , we have the possibility of introducing a `time difference operator', $\bar\tau$ , which we will call the duron. This object satisfies the commutator relations
\begin{eqnarray*}
[\bar T,\bar\epsilon]=[\bar\tau,\bar E]=i
\end{eqnarray*}
and
\begin{eqnarray}
[\bar T, \bar E]=[\bar\tau,\bar\epsilon]=[\bar T,\bar\tau]=[\bar E,\bar\epsilon]=0
\end{eqnarray}									
where we have written $\bar\epsilon$  for $\bar L$  to bring out the symmetry. Hiley and Fernandes \cite{bhmf97} have already suggested these relationships in the context of finding `operators' for time.  In particular they interpreted $\bar\tau$  as the mean time spent passing between two energy states.  Here we will suggest a different interpretation.
\newpage

\section{Bi-algebras and the Bogoliubov transformations}.

Before discussing the meaning of  $\bar\tau$ in more detail let me return to my way of thinking about the bi-algebra. I have proposed that the evolution of a quantum process does not proceed at an instant of time at a point in space, but through the ambiguous region of phase space that I have called a `moment'.  We consider the relation between the two sides of this moment, describing one side as information coming from the past while the other side is to do with the possible developments for the future.  

I have spoken at times rather dramatically about this latter feature as `information coming from the future'.  But such a way of talking is not that outrageous that it has not been suggested before.  For example Cramer \cite{jc86}  in his transactional interpretation of quantum mechanics uses the advanced potentials to carry information from the future.  The transaction is a `handshake' between emitter and the absorber participants of a quantum event.  This notion, in turn, has a resonance with an earlier proposal of Lewis \cite{gl26a}, \cite{gl26b} who has based his thinking on the following idea.  In the rest frame of a photon time dilation suggests that there is no time lapse between emission and absorption and because of the length contraction, there is no distance between the emitter and absorber either.  The light ray is a primary contact between the two ends of the process. These are both very radical ideas and unfortunately I have never known what to make of them so I have introduced the notion of a `moment' hoping that $\Delta t$, when projected into a space-time frame is small, but as these two examples show this may be a too conservative view to adopt! 

Recently I was very happy to meet with Giuseppe Vitiello  to discuss some of his extremely interesting ideas on dissipative quantum systems.  His ideas are, perhaps, even more conservative and therefore probably more reliable, yet they seem to fit into the overall scheme I am discussing here.  His work is reported in a series of papers in Vitiello \cite{gv95}, Celeghini, Rasetti and Vitiello \cite{crv92}, Celeghini {\em et al} \cite{ce98} and Iorio and Vitiello \cite{iv95}. I will rely heavily on the mathematics contained in these papers. 

They are interested in quantum dissipation, which they explore in terms of a pair of coupled dissipative oscillators, one emitting energy, the other absorbing energy.  In terms of our two-sided evolution discussed above, we find one `side' of the process is seen as representing the system while the other `side' is seen as representing the environment, the latter acting as a sink for the dissipated energy.  

In this model the degrees of freedom of the system are described by a set of annihilation operators $\{a_{k}\}$, while the environment is described by the set $\{\tilde a_{k}\}$ .  Thus there is a doubling of the mathematical structure.  The extra field variables   describing the `environment' are a mirror image of the variables used to describe the system.  Not only is a spatial mirror image but it is also a `{\em time-reversed} mirror image' as Vitiello \cite{gv96} puts it.  So the `environment sink' appears to be acting as if it were `anticipating the future'.

Let us leave the imagery for the moment and move on to see how the ideas work mathematically.  For this we will need to introduce some more formalism. So far we have introduced elements of our bi-algebra by effectively defining two sets of co-products which we will now express formally as
\begin{eqnarray}
\Delta_{+}\bar A=\bar A\otimes 1+ 1 \otimes \bar A\hspace{0.5cm}\mbox{and}\hspace{0.5cm}
\Delta_{-}\bar A=\bar A\otimes 1- 1 \otimes \bar A
\end{eqnarray}								
We have then shown that when we go to the limit $\Delta t\rightarrow 0$, we produce two dynamical equations, namely,
\begin{eqnarray}
i\frac{\partial \bar\rho_{V}}{\partial t}+\bar L\bar\rho_{V}=0\quad\mbox{and}\quad
\lim_{\Delta t\rightarrow 0}\left(2i\frac{\partial \bar\rho_{V}}{\partial t}\right)+\bar H_{+}\bar\rho_{V}=0
\end{eqnarray}								
But what do we make of the general co-products and the commutation relations listed in equations (\ref{eq:BP1})-(\ref{eq:BC})?  To explore these let us first make a Bargmann transformation from the Heisenberg algebra to the boson algebra of annihilation and creation operators.  This will enable us to immediately relate our work to that of Vitiello \cite{gv95}  and Celeghini {\em et al} \cite{ce98}.  Thus writing
\begin{eqnarray*}
&a=\bar x_{1}+i\bar p_{1}\quad\quad&\tilde a=\bar x_{2}+i\bar p_{2}\\
&a^{\dag}=\bar x_{1}-i\bar p_{1}\quad\quad&\tilde a^{\dag}=\bar x_{2}-i\bar p_{2}
\end{eqnarray*}
We can immediately make contact with equation (\ref{eq:U2H}) by using the well-known generator of the Bogoliubov transformation 
\begin{eqnarray}
G=-i(a^{\dag}\tilde a^{\dag}-a\tilde a)	\label{eq:BG}
\end{eqnarray}								
Then applying this to the vacuum state  $|0,0\rangle$, we find a new vacuum state $|0(\theta)\rangle$  given by
\begin{eqnarray}
|0(\theta)\rangle=\exp(i\theta G)|0,0\rangle=\sum_{n}c_{n}(\theta)|n\rangle\otimes|n\rangle	\label{eq:VACTH}
\end{eqnarray}								
This means that by doubling the algebra we can immediately see the similarity with equation (\ref{eq:U2H}) and this opens up the possibility of linking thermodynamics and quantum phenomena in a direct way, which is different from the thermal ensemble methods used in Bose-Einstein and Fermi statistics.  Doubling the algebra means doubling the degrees of freedom, so that we have a new process in addition to the usual dynamics.  

Umezawa \cite{hu93} gives a detailed discussion of a possible way of understanding this extra degree of freedom.  We will not discuss his ideas here, but suggest another way of exploiting these extra degrees of freedom to proved a better understanding of the notion of time.  To bring this possibility out  let us first go deeper and develop the boson bi-algebra a bit further by defining the following co-products based on equations (\ref{eq:BP1}) and (\ref{eq:BP2}),
\begin{eqnarray}
\Delta_{+}a=a\otimes 1+ 1 \otimes a=a+\tilde a;\quad\quad
\Delta_{-}a=a\otimes 1- 1 \otimes a=a-\tilde a.	\label{eq:DA}
\end{eqnarray}								
\begin{eqnarray}
\Delta_{+}a^{\dag}=a^{\dag}\otimes 1+ 1 \otimes a^{\dag}=a^{\dag}+\tilde a^{\dag};\quad\Delta_{-}a^{\dag}=a^{\dag}\otimes 1- 1 \otimes a^{\dag}=a^{\dag}-\tilde a^{\dag}.  \label{eq:DAD}
\end{eqnarray}								
We see immediately that these co-products are identical to those introduced by Celeghini et al \cite{ce98} but we can go further and form
\begin{eqnarray}
A=\tfrac{1}{\surd 2}(a+\tilde a)={\surd 2}(\bar X+i\bar P);\quad
A^{\dag}=\tfrac{1}{\surd 2}(a^{\dag}+\tilde a^{\dag})={\surd 2}(\bar X-i\bar P)		\label{eq:AAD}
\end{eqnarray}								
and
\begin{eqnarray}
B=\tfrac{1}{\surd 2}(a-\tilde a)=-{\surd 2}(\bar\eta+i\bar P);\quad
B^{\dag}=\tfrac{1}{\surd 2}(a^{\dag}-\tilde a^{\dag})=-{\surd 2}(\bar\eta-i\bar \pi)		\label{eq:BBD}
\end{eqnarray}								
These operators lie at the heart of their approach.  In our approach we see that these operators have a very simple interpretation.  They are simply the annihilation and creation operators of the mean position variables and the difference variables respectively. Thus
\begin{eqnarray*}
\bar X=\tfrac{1}{\surd 8}(A+A^{\dag})\quad\mbox{and}\quad
\bar P=\tfrac{i}{\surd 8}(A-A^{\dag})\\
\bar \eta=\tfrac{1}{\surd 2}(B+B^{\dag})\quad\mbox{and}\quad
\bar \pi=\tfrac{i}{\surd 2}(B-B^{\dag})
\end{eqnarray*}
In other words the operators $A$ and $B$ are the algebraic way of defining the ambiguous moments of in our algebraic phase space.  They are the variables that we need to describe the unfolding process that forms the basis of our paper.

Now I want to follow Celeghini et al \cite{ce98} further and generalise our approach by deforming the bi-algebra.  We do this by defining the co-product
\begin{eqnarray}
\Delta_{+}a_{q}=a_{q}\otimes q+q^{-1}\otimes a_{q}\hspace{1cm}
\Delta_{+}a^{\dag}_{q}=a^{\dag}_{q}\otimes q+q^{-1}\otimes a^{\dag}_{q}
\end{eqnarray}								
where we will write $q=e^{\theta}$ where $\theta$ is some parameter, the physical meaning of which has yet to be determined.  Then
\begin{eqnarray}
A_{q}=\frac{\Delta a_{q}}{\surd [2]_{q}}=\tfrac{1}{\surd[2]_{q}}(e^{\theta}a+e^{-\theta}\tilde a);\quad
B_{q}=\tfrac{1}{\surd[2]_{q}}\frac{\delta}{\delta\theta}\Delta a_{q}=\tfrac{1}{\surd[2]_{q}}(e^{\theta}a+e^{-\theta}\tilde a)\nonumber\\
+h.c.\hspace{5cm}
\end{eqnarray}								
The $A_{q}$ and $B_{q}$ are then the deformed equivalents of equations (\ref{eq:AAD}) and (\ref{eq:BBD}). Notice also that
\begin{eqnarray}
\Delta_{-}A_{\theta}=\frac{\delta}{\delta\theta}\Delta_{+}A_{\theta}
\quad\mbox{and}\quad\Delta_{-}A=\lim_{\theta\rightarrow 0}\frac{\delta}{\delta\theta}\Delta_{+}A_{\theta}.
\end{eqnarray}								
so that the two sets of co-products defined in equations (\ref{eq:DA}) and (\ref{eq:DAD}) are not independent. With these definitions it is not difficult to show that we can write
\begin{eqnarray}
A(\theta)=\tfrac{1}{\surd 2}(a(\theta)+\tilde a(\theta))\quad\mbox{and}\quad B(\theta)=\tfrac{1}{\surd 2}(a(\theta)-\tilde a(\theta))  \label{eq:ABTH}
\end{eqnarray}								
So that
\begin{eqnarray}
a(\theta)=\tfrac{1}{\surd 2}(A(\theta)+B(\theta))=a\cosh\theta -\tilde a^{\dag}\sinh\theta
\end{eqnarray}								
and
\begin{eqnarray}
\tilde a(\theta)=\tfrac{1}{\surd 2}(A(\theta)-B(\theta))=\tilde a\cosh\theta -a^{\dag}\sinh\theta
\end{eqnarray}								
This is immediately recognised as the Bogoliubov transformation from the set $\{a,\tilde a\}$ of annihilation and creation operators  to a new set $\{a(\theta),\tilde a(\theta)\}$.  This result justifies the use of the Bogoliubov generator given in equation (\ref{eq:BG}), which was used to construct the GNS ket given in equation  (\ref{eq:VACTH}).

\section{Unfolding through inequivalent representations?}

Having put a formalism in place, I now want to consider how all this leads to a radically new way of looking at the way quantum processes unfold in time.  My ideas go back to the early eighties when David Bohm and I were discussing how we could think about the type of process underlying quantum phenomenon.  Most of this work was unpublished essentially because I did not have an adequate understanding of the mathematics needed.  However Bohm \cite{db86} did publish some of the background relevant to the ideas I am developing here. There perhaps for the first time he makes a clear statement as to what we were thinking.  I quote 
\begin{quote}
All these relationships (of moments of enfoldment) have to be understood primarily as being between the implicate ``counterparts" of these explicate moments.  That is to say, we no longer suppose that space-time is primarily an arena and that the laws describe necessary relationships in the development of events as they succeed each other in this arena.  Rather, each law is a structure that interpenetrates and pervades the totality of the implicate order.
\end{quote}

Implicit in this was the idea that space-time itself would emerge at some higher explicit level (Hiley \cite{bjh91}).  All of this early discussion could easily  be dismissed as `somewhat vague', but we did try to make it more specific by arguing that the inequivalent representations contained within quantum field theory would play a key role.  However we could not see how to make the mathematics work. 

In the general context of Bohm's ideas, the vacuum state should not be regarded as absolute and self-contained.  Rather each vacuum state provides the basis for what we called an explicate order so that a set of inequivalent vacuum states could be thought of as providing an array of explicate orders, all embedded in the overall implicate order in which all movement is assumed to take place. The movement between inequivalent representations, between inequivalent vacuum states, is then regarded as a movement from one explicate order to another. 

This movement, as we have seen, is described by a Bogoliubov transformation and should be distinguished from the unfolding-enfolding transformation that Bohm describes with the metaphor of the jar of glycerine demonstration \cite{db80}.  Mathematically this is just an automorphism of the kind $e'=MeM^{-1}$ as was discussed in \cite{bh08}.  Within this structure we found the explanation as to why in a single Hilbert space formalism nothing {\em actually} happens. The inner automorphisms of the algebra of operators are simply a re-description of the {\em potentialities} of the process so that every unitary transformation becomes merely a re-expression of the order.  In this sense everything is a {\em potentiality}.  

But what about the actual occasions?  This has been the continuing difficulty of the `measurement problem'.  Where do the actual events arise in the quantum formalism?  First we should notice that in quantum field theory, the vacuum kets $|0(\theta)\rangle$  belong to inequivalent representations of the boson algebra.  Our suggestion is that not only is there a movement within each inequivalent representation but there is also another movement involved and this is the movement {\em between} inequivalent representations and thus between these inequivalent vacuum states.  The key question how is this movement described mathematically.

The answer appears to lie in the relationship between the two co-products described by equation (\ref{eq:ABTH}) as Celeghini {\em et al} \cite{ce98} have already pointed out.  It is this feature that allows us to discuss the movement between inequivalent representations.  To explain this idea let us define
\begin{eqnarray}
p_{\theta}=-i\frac{\delta}{\delta \theta}.
\end{eqnarray}								
We can then think of $p_{\theta}$ as a conjugate momentum to the internal degree of freedom $\theta$ so that this momentum can be thought of as describing the movement between inequivalent Hilbert spaces.  This identification becomes even more compelling once we realise that
\begin{eqnarray}
-i\frac{\delta}{\delta \theta}a(\theta)=[G,a(\theta)]\quad\mbox{and}\quad
-i\frac{\delta}{\delta \theta}\tilde a(\theta)=[G,\tilde a(\theta)]
\end{eqnarray}								
Here $G$ is the generator of the Bogoliubov transformation given in equation (\ref{eq:BG}).  Indeed if we use this generator then for a fixed value of $\bar\theta$  we have
\begin{eqnarray}
\exp(i\bar\theta p_{\theta})a(\theta)=\exp(i\bar\theta G)a(\theta)\exp(-i\bar\theta G)=a(\theta +\bar\theta).
\end{eqnarray}								
which is equivalent to the transformation from $|0(\theta)\rangle\rightarrow|0(\theta+\bar\theta)\rangle$.  Furthermore and even more importantly from our point of view the movement is expressed in terms of an inner automorphism of the algebra\footnote{The inner automorphism is a way of expressing the enfolding and unfolding movement.}.

\subsection{The Role of Time}

Finally I want to return specifically to the question of time.  In the bi-algebra we have two elements of time,
\begin{eqnarray}
\bar T=\Delta_{+}t=t\otimes 1+1 \otimes t\quad\mbox{and}\quad \tau=\Delta_{-}t=t\otimes 1 - 1 \otimes t.
\end{eqnarray}									

Since  $\Delta_{+}t$ and $\Delta_{-}t$  are related, $T$ and $\tau$ are not independent.  If we regard $\psi_\theta(x,t)$  as the eigenfunction of $\bar T$ so that $\bar T\psi_\theta(x,t)=t\psi_\theta(x,t)$ then we will represent $\tau$ by $-i\partial/\partial t$.  In the conjugate representation $\phi_\theta(x,\tau)$ is the eigenfunction of  $\tau$, then $\bar T$ will take the form $i\partial/\partial t$.  Here I am merely exploiting the analogy between the $x$- and the $p$-representations where the operators are $(x,-i\partial/\partial x)$  and $(p,i\partial/\partial p)$ respectively.  

How are we to understand this structure?  When $\bar T$ is diagonal, we remain within a single Hilbert spaces parameterised by $\theta$.  Its eigenvalue, $t$, will then be the Schr\"{o}dinger time.  This means the potentialities are changing with time although no irreversible process is taking place.  The system remains within this Hilbert space, getting older as it were but not actualising.  Bohm \cite{db87} calls $\bar T$ the implication parameter and regards it as a measure of the age of the system.

Our proposal is that an actual change comes about when a transformation to a different inequivalent vacuum state occurs or, in other words, to a new Hilbert space.  Notice that during this transformation, $\bar T$ is no longer diagonal implying that Schr\"{o}dinger time is ambiguous during the transition process.  Thus the Schr\"{o}dinger equation is no longer valid.

  A new process unfolds and $\tau$ becomes diagonal.  This means that the time between inequivalent states is well defined signifying a Bogoliulov transformation is taking place.  This would then tie in with the idea of Hiley and Fernandes \cite{bhmf97}, where they regarded $\tau$ as a measure of the time between states, but in this paper it is regarded as a measure of time between inequivalent vacuum states.  The fact that $\theta$ and its conjugate $p_{\theta}$ do not commute implies that transition between inequivalent states is not sharp and requires just the kind of ambiguity we have suggest accompanies the notion of a moment.

This kind of ambiguity is not surprising as quantum theory already tells us that energy and time are complementary variables.  So why do we insist on the evolution of a process with a definite energy occurring at a definite instant of time?  Mean energy can be conserved but surely to have change, we must have some ambiguity in each moment of time to allow for the creation of a new structure.   Here we are exploiting the tension between what has gone with what is to come. We must have a break between the structure that has been and the new structure that is to come.  This implies that are many coexisting instants of time with various weightings in the same moment.  In this way not only does quantum theory contain spatial non-locality but that it also contains a `non-locality' in time as proposed by Peres \cite{ap80}.

This discussion suggests a very different view of time evolution.  It is not a substitution of one point-like event evolving into another infinitesimal later point-like event.  Rather there is a enfolding-unfolding of an extended structure as has been suggested by Bohm \cite{db86} when he writes
\begin{quote}
Becoming is not merely a relationship of the present to a past that is gone. Rather, it is a relationship of enfoldments that actually are together in the present moment. Becoming is an actuality.
\end{quote}

In Umezawa \cite{hu93} the parameter $\theta$ is asociated with temperature.  Indeed it is tempting to regard $\theta$ as the inverse of $\beta$, i.e. $\theta$ is proportional to the temperature.  However I am reluctant to make this a definitive step at this stage because I am very aware of the idea of modular flow introduced by Rovelli \cite{cr93} and Connes and Rovelli \cite{accr94} which has some direct relevance to what I am discussing here.  These papers have an extensive discussion on the thermodynamic origin of time.  They have probed deeper into the mathematical structure implicit in the work I am discussing and have shown how the Tomita-Takesaki theorem provides this connection between time and the thermal evolution of a quantum system.  There are clearly connections between this work and the tentative proposals I have outlined in my paper. There is much more to be said but this must be left for another publication.

\section{Acknowledgements}.

I would like to thank the members of the TPRU for their invaluable help in trying to straighten out the ideas expressed in this paper.  I would like to thank Albrecht von M\"{u}ller and Thomas Filk for inviting me to participate in their stimulating meeting at the Parmenides Foundation in Munich.  


\end{document}